\documentclass[twoside,twocolumn,10pt]{article}

\usepackage[sc]{mathpazo} 
\usepackage[T1]{fontenc} 
\usepackage{bold-extra}
\usepackage{microtype} 

\usepackage[english]{babel} 

\usepackage[margin=2cm]{geometry} 
\usepackage[small,labelfont=bf,up,textfont=bf]{caption} 
\usepackage{booktabs} 

\usepackage{lettrine} 

\usepackage{enumitem} 
\setlist[itemize]{noitemsep} 

\usepackage{abstract} 

\usepackage{titlesec} 
\renewcommand\thesection{\arabic{section}} 
\renewcommand\thesubsection{\arabic{subsection}} 
\titleformat{\section}[block]{\large\bfseries\centering}{\thesection.}{1em}{} 
\titleformat{\subsection}[block]{\normalsize\bfseries}{\thesection.\thesubsection.}{.5em}{} 

\usepackage{fancyhdr} 
\usepackage{titling} 
\usepackage[numbers]{natbib}

\usepackage[unicode=true,bookmarks=false,pdfstartview={FitH},colorlinks=true,citecolor=blue,linkcolor=blue,urlcolor=blue]{hyperref} 

\usepackage{amsmath,amsfonts,amsthm}
\usepackage{graphicx}
\usepackage{epstopdf}
\usepackage{algorithmic}
\usepackage{amsopn}
\usepackage{bm}
\usepackage{caption}
\usepackage{float}
\usepackage{subcaption}

\ifpdf
  \DeclareGraphicsExtensions{.eps,.pdf,.png,.jpg}
\else
  \DeclareGraphicsExtensions{.eps}
\fi

\newcommand{\citeeg}[1]{(e.g.,~\cite{#1})}
\newcommand{\as}{\quad\textrm{as}\quad}
\newcommand{\N}{\mathcal{N}}
\renewcommand{\P}{P}
\renewcommand{\Pr}{\mathrm{Pr}}
\DeclareMathOperator{\var}{var}
\DeclareMathOperator{\cov}{cov}

\newcommand{\B}{\mathrm{Bernoulli}}
\newcommand{\I}{\mathbb I}
\newcommand{\cc}{\bm{c}}
\newcommand{\y}{\bm{y}}

\newcommand{\E}{\mathbb E}
\newcommand{\Dto}{\stackrel {D}{\longrightarrow}}
\makeatletter
\newcommand{\iid}{\stackrel {\operator@font{iid}}{\sim}}
\newcommand{\ind}{\stackrel {\operator@font{ind}}{\sim}}
\makeatother
\newcommand{\btheta}{\bm{\theta}}
\newcommand{\ii}{\delta}
\newcommand{\g}{{\bm{g}}}
\newcommand{\s}{\bm{s}}
\renewcommand{\SS}{\bm{\Sigma}}

\newcommand{\cp}{\rho}

\newcommand{\SM}{S_M}
\newcommand{\II}{{\mathcal S}}
\newcommand*\ud{\mathop{}\!\mathrm{d}}
\newcommand{\nobs}{N}
\newcommand{\nix}{i}
\DeclareMathOperator*{\argmax}{arg\,max}

\theoremstyle{plain}
\newtheorem{theorem}{Theorem}[section]
\newtheorem{corollary}[theorem]{Corollary}
\pretitle{\begin{center}\Large\bfseries} 
\posttitle{\end{center}} 
\title{A Convergence Diagnostic for Bayesian Clustering\thanks{This research was partly supported by Natural Sciences and
Engineering Council of Canada (NSERC) grants RGPIN-217398-13 (Asgharian), RGPIN-2014-04255 (Lysy), and RGPIN-418034-12 (Partovi Nia).  All authors contributed equally to this work.}} 
\author{%
  \textsc{Masoud Asgharian}
  \\[1ex] 
\small Department of Mathematics and Statistics \\
\small McGill University 
\and 
\textsc{Martin Lysy} \\[1ex] 
\small Department of Statistics and Actuarial Science \\ 
\small University of Waterloo 
\and
\textsc{Vahid Partovi Nia} \\[1ex]
\small Noah's Ark Research Lab \\
\small Huawei Technologies Canada
}
\date{June 12, 2019} 
\renewcommand{\maketitlehookd}{%
\begin{abstract}
  \noindent{} In many applications of Bayesian clustering, posterior sampling on the discrete state space of cluster allocations is achieved via Markov chain Monte Carlo (MCMC) techniques.  As it is typically challenging to design transition kernels to explore this state space efficiently, MCMC convergence diagnostics for clustering applications is especially important.  For general MCMC problems, state-of-the-art convergence diagnostics involve comparisons across multiple chains.  However, single-chain alternatives can be appealing for computationally intensive and slowly-mixing MCMC, as is typically the case for Bayesian clustering.  Thus, we propose here a single-chain convergence diagnostic specifically tailored to discrete-space MCMC.  Namely, we consider a Hotelling-type statistic on the highest probability states, and use regenerative sampling theory to derive its equilibrium distribution. 
  By leveraging information from the unnormalized posterior, our diagnostic protects against seemingly convergent chains in which the relative frequency of visited states is incorrect. 
  The methodology is illustrated with a Bayesian clustering analysis of genetic mutants of the flowering plant \emph{Arabidopsis thaliana}.

  \smallskip

  \noindent{}\emph{Keywords:} Bayesian clustering, Markov chain Monte Carlo, convergence diagnostic, Hotelling statistic, regenerative sampling.
\end{abstract}
}

\begin{document}

\maketitle


\section{Introduction}
\label{section:intro}

Clustering may be described as the task of partitioning data into homogeneous groups.  While classical clustering techniques employ geometric measures of dissimilarity to distinguish between groups~\citep*{Hartigan:Clustering:1975}, modern approaches are based on probabilistic models where homogeneous groups of data follow the same distribution~\citep*{Murua:Stanberry:Stutzle:2008,Everitt:Landau:Leese:Stahl:Cluster:2011}. From the perspective of statistical inference, probabilistic clustering may be regarded as fitting a mixture model with the number of components unknown.

When the number of components is fixed, observations can be readily allocated to clusters by maximum likelihood via the EM algorithm. Subsequently, the number of clusters often is determined by model selection criteria such as AIC and BIC~\citep*{Fraley:Raftery:Model-Based:2002}. In Bayesian model-based clustering, a prior distribution is assumed on both parameters and groupings~\citep*{Heard:Holmes:Stephens:Mosquitoes:2006}, such that the posterior distribution is on all possible allocations of the $\nobs$ observations to $C$ clusters, $1 \le C \le \nobs$.  When only the maximum \emph{a posteriori} (MAP) allocation of the data is sought, deterministic search algorithms such as Bayesian hierarchical clustering~\citep*{Heller:Ghahramani:Bayesian:2005} may be used.  Alternatively, consensus clustering~\citep*[e.g.,][]{strehl.ghosh02,monti.etal03} attempts to aggregate multiple cluster allocations, often leading to superior partitioning of the data~\citep*{topchy.etal03}.  In the Bayesian setting, the clusters to be aggregated are typically sampled from their posterior distribution using Markov chain Monte Carlo (MCMC) techniques~\citep*[e.g.,][]{Liu:2001, Robert:Casella:MonteCarlo:2004}.

While generic MCMC for Bayesian clustering is fairly straighforward, designing efficient transition kernels is a challenging task. 
For one thing, even for small $\nobs$, the cardinality of the space of all clusters -- denoted by the Bell number $B(\nobs)$ -- is monumentally large.  With only $\nobs = 14$ (as in the upcoming application), we have $B(14) \approx 1.9\times 10^8$. 
For $\nobs = 100$ observations, $B(100) \approx 4.8\times 10^{115}$. 
Furthermore, most transition kernels on the state space of clusters reallocate a single observation at a time~\citeeg{neal00}, which tends to result in very slow MCMC convergence.  More sophisticated kernels reallocating groups of observations include split-merge proposals~\citep{Jain:Neal:SpiltMerge:2004,Jain:Neal:2007} and reversible-jump MCMC~\citep{green95,richardson.green97,green.richardson01}.  However, carefully-tuned interweaving between singleton and group proposals is needed to achieve good MCMC mixing~\cite{Jain:Neal:SpiltMerge:2004}, suggesting that convergence diagnostics in Bayesian clustering are especially important to monitor. 


For MCMC with mixture models,~\cite{Brooks:Giudici:Philippe:2003} propose a nonparametric convergence criterion based on Markov chain subsampling.  However, subsampling estimators can suffer from a considerable loss of efficiency~\citep{Geyer:1992, MacEachern:Berliner:1994}.  For general MCMC, a simple and versatile convergence diagnostic is that of~\cite{Jones:Haran:Caffo:Neath:2006}, which computes the ratio between selected MCMC sample moments and their standard errors, with various methods having been proposed to calculate the latter~\citeeg{Jones:Haran:Caffo:Neath:2006,flegal.jones10,liu.flegal18,vats.flegal18}.  
Tolerance levels on the coefficient of variation (CV) are then used to assess convergence.  However, this approach can fail when the MCMC becomes trapped in a local mode.  In this case, prior to sufficient mixing of the chain, sample moments seemingly converge but to the wrong value.  To overcome this issue, a widely-used diagnostic measure is that of~\cite{gelman.rubin92} and its variants~\citeeg{brooks1998general,vats.knudson18}, wherein multiple chains are run from overdispersed starting points and between-chain and within-chain variances are compared.  While these methods are much more effective in detecting local modes, running multiple chains can be statistically inefficient compared to running a single chain for the same amount of time~\citep{Cowles:Carlin:Review:1996}.  The problem is particularly severe for slowly mixing algorithms, as is typically the case for Bayesian clustering.

In this paper, we propose a single-chain convergence diagnostic specifically tailored to discrete-space MCMC.  Namely, we consider a Hotelling-type statistic on the relative frequency of visited states.  While no convergence diagnostic can positively ascertain that a given MCMC sample is representative of its equilibrium distribution~\citep{Cowles:Carlin:Review:1996}, ours can detect seemingly convergent chains in which the relative frequency of visited states is incorrect.  This is done in the spirit of~\cite{ZellnerMin:Gibbs:1995} by leveraging information from the unnormalized equilibrium distribution.  For clustering applications, this is available for mixtures of exponential families with conjugate priors, for which the model parameters can be integrated out.  
Our diagnostic quantifies lack of convergence via tail probabilities of its asymptotic distribution, which we derive by extending the regenerative sampling Central Limit Theorem~\citep{Mykland:Tierney:Yu:Regeneration:1995,Hobert:Jones:Presnell:Rosenthal:2002,Jones:Haran:Caffo:Neath:2006} to a multivariate setting.  A related approach is that of~\cite{johnson98}, but for which running an additional coupling chain is required.

The remainder of this paper is organized as follows. Section~\ref{section:clustering_and_MCMC} describes the general framework of Bayesian clustering in which our convergence diagnostic may be applied.  
In Section~\ref{section:conv} we define our convergence statistic and derive its asymptotic distribution.  
In Section~\ref{section:application}, we illustrate the benefits of our methodology in a Bayesian clustering analysis of genetic mutants of the flowering plant \emph{Arabidopsis thaliana}.  We conclude in Section~\ref{section:discuss} with potential directions for future work.


\section{Bayesian Clustering}
\label{section:clustering_and_MCMC}

In Bayesian clustering, each observation has a corresponding unknown grouping parameter which assigns it to a specific cluster.  Let $\y = \{y_{\nix}\}_{\nix=1}^\nobs$ represent the observations and $\cc = \{c_\nix\}_{\nix=1}^\nobs$ the unknown grouping parameters called labels, i.e., $c_\nix = c \in \{1,\ldots,C\}$ if $y_\nix$ is allocated to cluster $c$.  In order to impose uniqueness in cluster labeling, we assume that the grouping parameters are in increasing order, i.e., the first observation, $y_1$, always has label $1$; the second observation has label $1$ if it belongs to the same group as $y_1$; otherwise, it has label $2$, and so forth. Furthermore, we assume that there are no empty clusters.  The likelihood function is then given by
\[
p(\y \mid \btheta, \cc) = \prod_{c=1}^C \prod_{t: c_t = c} p(y_t \mid \btheta, c),
\]
where $\btheta$ are the unknown model parameters. We assume, conditional on $\cc$ and $\btheta$, that the observations are independent within and across clusters, which is called a \emph{partition model}~\citep{Hartigan:PartitionModels:1990}. Since the goal is to estimate the grouping parameter $\cc$, the ideal scenario involves fitting a model with closed-form marginal posterior distributions~\citep{Heller:Ghahramani:Bayesian:2005, Heard:Holmes:Stephens:Mosquitoes:2006}. In other words, the model parameters are integrated out with respect to their prior distribution given $\cc$:
\begin{equation}\label{rao}
p(\y\mid \cc)=\int \left\{\prod_{c=1}^C\prod_{t: c_t=c} p(y_t\mid
\btheta, \cc)\right\} \pi(\btheta \mid \cc) \ud\btheta.
\end{equation}
A large class of models for which $p(\y\mid\cc)$ is available in closed form are exponential families with conjugate priors, of which we give an example in Section~\ref{section:application}. The state space of interest is that of all possible allocations under the posterior distribution
$p(\cc \mid \y) \propto p(\y \mid \cc) \pi(\cc)$,
where $\pi(\cc)$ is the prior distribution on allocations. The Rao-Blackwellization of~\eqref{rao} reduces
the variance of MCMC-based estimators and facilitates the exploration of $p(\cc \mid \y)$ by MCMC. The current literature offers several choices for the prior distribution $\pi(\cc)$~\citeeg{McCullagh:Yang:2006,Heard:Holmes:Stephens:Mosquitoes:2006,Booth:Casella:Hobert:Clustering:2008}.

\section{Convergence Diagnostic}\label{section:conv}

\subsection{Preliminaries}

Let $\{X_t\}_{t \ge 1}$ be an irreducible, aperiodic Markov chain with discrete state space $\SM$ of cardinality $M$.
In the context of Bayesian clustering for $N$ observations, 
$X_t$ is an integer which identifies a distinct grouping $\cc$, and $M = B(N)$ is the Bell number.  We therefore use ``state'' and ``grouping'' interchangeably from this point.
Let $\bm P= [ \P_{ij}]_{1\le i,j \le M}$ denote the transition probability matrix for the Markov chain. By the ergodic theorem \citep{Meyn:Tweedie:1993}, there exists a unique stationary distribution $\bm{\Pi} = (\Pi_1, \ldots, \Pi_M)$, such that $\bm{\Pi} \bm P = \bm{\Pi}$, satisfying $\Pi_j = \lim_{k\rightarrow \infty} \P^{(k)}_{ij}, \forall i,j \in \SM$, where $\P^{(k)}_{ij}$ is the transition probability from state $i$ to state $j$ in $k$ steps.

Now suppose that $\bm{\Pi}$ is known up to a normalizing constant.  That is, we know
\[
\pi_i = Z \Pi_i \quad \forall\, i \in \SM,
\]
where $Z = \sum_{i=1}^M \pi_i > 0$.  We assume that the state space $\SM$ is prohibitively large, such that enumerating all states to compute the normalizing constant is computationally infeasible.  This is the setting for model-based Bayesian clustering, when the model parameters can be integrated out as in~\eqref{rao}.

In order to define our convergence statistic and its asymptotic distribution, we employ the technique of \emph{regenerative sampling}~\citep{Mykland:Tierney:Yu:Regeneration:1995}. Suppose that the Markov chain has been run for $n$ iterations.  For any fixed state $\ii \in S_{M}$, let $\tau_{r}$ be the $(r+1)$th time $X_{t}$ visits state $\ii$, such that $X_{\tau_{r}}=\ii$. In other words, $\tau_{r}$ is the time of the $r$th return to state $\ii$ for $r>0$. Let $R=R(n)$ denote the number of returns to state $\ii$ --  or \emph{regeneration tours} -- in the $n$ Markov chain iterations. Since the Markov chain is aperiodic, it follows that $R \to \infty$ as $n \to \infty$.

Let $g(x)$ be a real-valued, $\bm \Pi$-integrable function on $\SM$.  The ergodic theorem implies that
\[
\bar{g}_{\tau_R}  = \frac{1}{\tau_R-1} \sum_{t=1}^{\tau_R-1} g(X_t) \quad \to \quad \E_{\bm\Pi} [g(X_t)] = \sum_{i \in \SM} g(i) \Pi_i
\]
with probability 1 as $R \to \infty$.  The quantity $\bar{g}_{\tau_R}$ is called the regenerative sampling (RS) estimator.  Note that $\tau_R$ is the start of the $(R+1)$st regeneration tour, hence the limits of the summation. It was shown by~\cite{Hobert:Jones:Presnell:Rosenthal:2002} that the Central Limit Theorem (CLT) also holds if $\{X_t\}$ is geometrically ergodic and $\E_{\bm\Pi} \big[|g(X_t)|^{2 + \epsilon}\big] < \infty$ for some $\epsilon > 0$, namely,
\begin{equation} \label{result:CLT}
\sqrt{R} \cdot ( \bar{g}_{\tau_R} - \mu_g ) \Dto \N(0, \sigma^2_g) \as R \to \infty,
\end{equation}
where $\mu_g = \E_{\bm\Pi} [g(X_t)]$ and $\sigma^2_g < \infty$. 
Furthermore,~\cite{Hobert:Jones:Presnell:Rosenthal:2002} go on to derive a consistent estimator of $\sigma^2_g$, and relate it to the familiar Markov chain CLT of~\cite{Chan:Geyer:1994}:
\begin{equation} \label{result:CLT2}
\sqrt{n} \cdot ( \bar g - \mu_g) \Dto \N(0, \gamma^2_g) \as n \to \infty,
\end{equation}
where $\bar g = n^{-1} \sum_{t=1}^n g(X_t)$,
\[
  \gamma^2_g = \var_{\bm\Pi} \big\{ g(X_t) \big\} + 2 \sum_{k=1}^{\infty} \cov_{\bm\Pi} \big\{ g(X_t), g(X_{t+k}) \big\} < \infty,
\]
and $\sigma_g^2 = \gamma_g^2 \Pi_\ii$.

\subsection{Diagnostic Tool}

Let $S_M = \coprod_{i=1}^{K+1} \II_i$ be a partition of the sample space.  Using the ergodic theorem, the RS estimator
\[
  \bar Q_i = \frac1{\tau_R-1} \sum_{t=1}^{\tau_R-1} \I(X_t \in \II_i)
\]
is a consistent estimator of $Q_i = \E_{\bm \Pi}[\I(X_t \in \II_i)] = Z^{-1}  q_i$, where  $q_i = \sum_{j \in \II_i} \pi_j$. Thus for large values of $R$, we expect the ratio $f_i = \bar Q_i / q_i$ to be close to $Z^{-1}$, for $1 \le i \le K$.  Hence, the $f_i$ are approximately constant when the Markov chain reaches equilibrium. Indeed, let $g_i(x) = \I(x \in \II_i)/q_i$ and $\g(x) = \big(g_1(x), \ldots, g_K(x)\big)$. Then by standard results in regenerative sampling theory~\citeeg{Mykland:Tierney:Yu:Regeneration:1995, Hobert:Jones:Presnell:Rosenthal:2002, Jones:Haran:Caffo:Neath:2006} we have
\begin{equation}\label{eq:gbar}
  \bar \g_{\tau_R} = \frac 1 {\tau_R - 1} \sum_{t=1}^{\tau_R-1} \g(X_t) = (f_1, \ldots, f_K) \to \bm 1_K \cdot Z^{-1},
\end{equation}
where $\bm 1_K = (1, \ldots, 1)$, and $R \cdot \var(\bar \g_{\tau_R}) \to \SS_\g$.  Thus, we consider the Hotelling statistic
\begin{equation}\label{eq:HT}
  T^2 = (\bar \g_{\tau_R} - \hat Z^{-1})'\SS_\g^{-1}(\bar \g_{\tau_R} - \hat Z^{-1}),
\end{equation}
where
\begin{align*}
  \hat Z^{-1} & = (\bm 1_K'\SS_\g^{-1}\bm 1_K)^{-1}\bm 1_K' \SS_\g^{-1} \bar \g_{\tau_R}\\
              & = \big(w_1(\SS_g), \ldots, w_K(\SS_g)\big)' \bar \g_{\tau_R} = \bm w(\SS_g)' \bar \g_{\tau_R}.
\end{align*}
Large values of $T^2$ indicate that the empirical probabilities $\bar Q_i$ are incorrectly weighted relative to each other;  the ratios $\bar Q_i/\bar Q_j$ are far from their true (known) values $Q_i/Q_j = q_i/q_j$, suggesting that the MCMC has not yet converged to its stationary distribution.  Our diagnostic tool quantifies large values of $T^2$ with respect to its asymptotic distribution, upon substituting the unknown variance $\SS_\g$ in~\eqref{eq:HT} with a consistent estimator.  Namely we have the following results.
\begin{theorem}\label{thm:main}
  Let $\g: \SM \to \mathbb R^K$ with $\g(x) = \big(g_1(x), \ldots, g_K(x)\big)$, and suppose there exists $\epsilon > 0$ such that $\E_{\bm\Pi} \big[|g_i(X_t)|^{2 + \epsilon}\big] < \infty$ for $i=1,\ldots,K$. Define $\bar \g_{\tau_R}$ as in~\eqref{eq:gbar}, and let $\s_r = \sum_{t=\tau_{r-1}}^{\tau_r-1} \g(X_t)$ denote the sums in each regeneration tour, $N_r = \tau_r - \tau_{r-1}$ the length of each tour, and $\bar N = R^{-1} \sum_{r=1}^R N_r$ the average tour length.  Then for an irreducible, aperiodic, discrete state space Markov chain with equilibrium distribution $\bm\Pi$,
  \begin{equation}\label{eq:regvar}
\hat \SS_{\tau_R} = \frac 1 {R \bar N^2} \sum_{r=1}^R (\s_r - N_r \bar \g_{\tau_R})(\s_r - N_r \bar \g_{\tau_R})'
\end{equation}
is a consistent estimator of $\SS_\g$.
\end{theorem}
As a consequence, we have the asymptotic distribution of a Hotelling-type regenerative sampling statistic:
\begin{corollary}\label{thm:hot}
  Let $g_i(x) = \I(x \in \II_i)/q_i$, $i = 1,\ldots,K$, and $\hat Z_{\tau_R}^{-1} = \bm w(\hat \SS_{\tau_R})' \bar \g_{\tau_R}$.
  Then the Hotelling-RS statistic
\begin{equation}\label{eq:HRS}
  T^2_{\tau_R} = (\bar \g_{\tau_R} - \hat Z_{\tau_R}^{-1})'\hat \SS_{\tau_R}^{-1}(\bar \g_{\tau_R} - \hat Z_{\tau_R}^{-1})
\end{equation}
asymptotically has a $\chi^2_{(K-1)}$ distribution.
\end{corollary}
The proofs of Theorem~\ref{thm:main} and Corollary~\ref{thm:hot} are given in Appendix~\ref{appendix:asym}.

\subsection{Practical Considerations}\label{sec:prac}

Suppose that the states are sorted by decreasing probability mass, $\Pi_1  \ge \cdots \ge \Pi_M$.  Then a simple choice for the regeneration tour counter $\ii$ and the partition sets $\II_i$ is
\[
  \ii = 1,  \qquad \II_i = i, \quad i = 1,\ldots,K.
\]
Thus, the Hotelling-RS statistic~\eqref{eq:HRS} focuses on the $K$ most probable states which are likely to dominate the analysis.  In practice, these high probability states are not known in advance.  However, they can be estimated from an MCMC sample by ranking the unnormalized probabilities $\pi_i = Z\Pi_i$ of all visited states.  Unfortunately, it is not possible to establish a lower bound on the quality of such an estimator.  That is, suppose that after $n$ steps the Markov chain has visited $M_n < M$ states, denoted by the set $S_n \subset \SM$.  Let $\bm P^\star_{1\le i,j\le M_n}$ denote the transition matrix restricted to $S_n$, of which the elements are
\[
  P^\star_{ij} = \Pr(X_{t+1} = j \mid \{X_{t} = i\}\cap\{j \in S_n\}).
\]
Then we can always find a transition matrix $\bm P$ on the whole space $\SM$ which is consistent with $\bm P^\star$, and for which $\Pr(X_t \in S_n) = \epsilon$ under the stationary distribution $\bm \Pi$.  This is achieved by taking a state $i \notin S_n$ to have very high probability $\Pi_i$, very high self-transition probability $p_{ii}$, and very low $p_{ji}$ for $j \in S_n$.  In this sense, the diagnostic tool only checks the relative frequencies between visited states.  

\section{Illustration}
\label{section:application}

\begin{figure*}[!htp]
\centering
\includegraphics[width=1.00\textwidth]{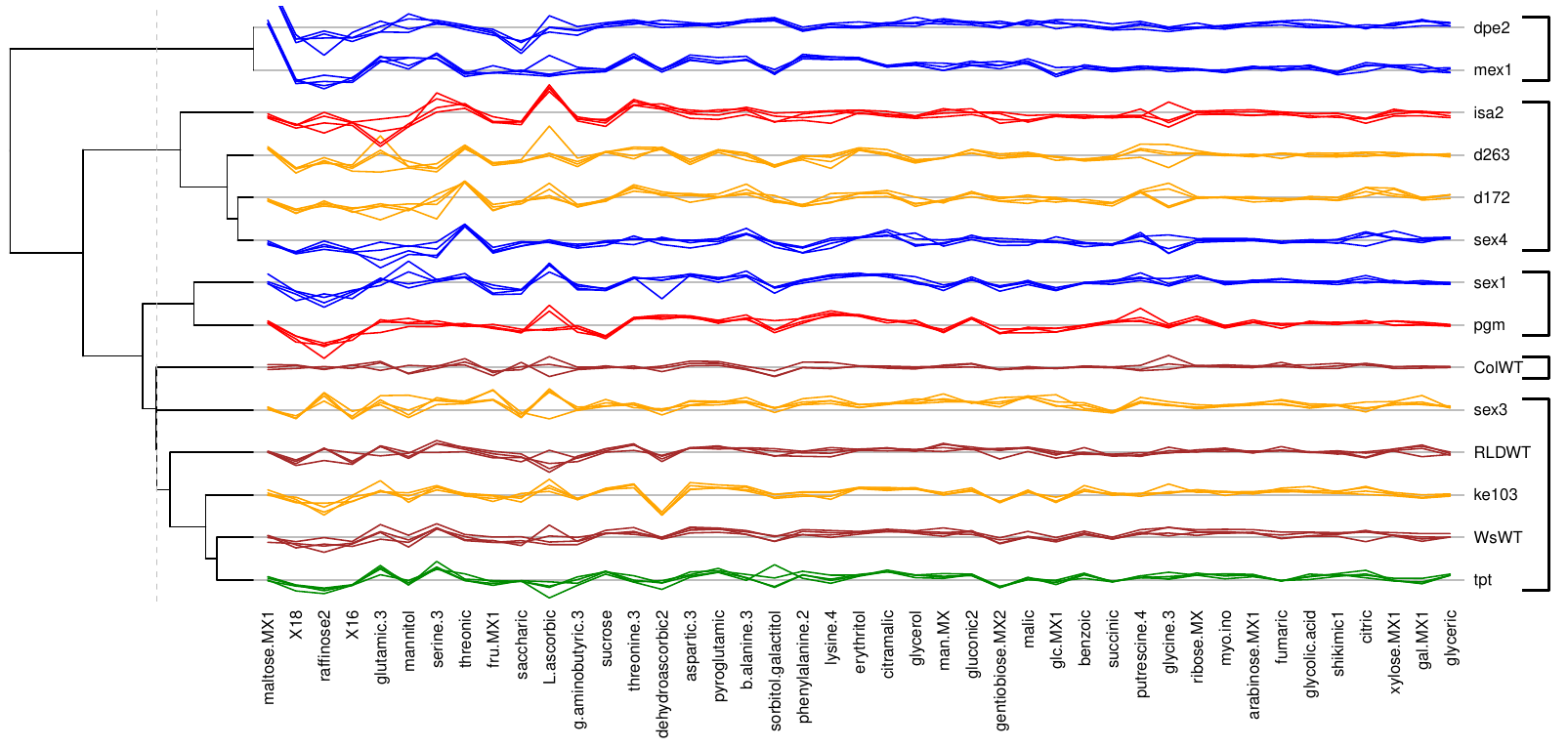}
\caption{Profile plot of metabolite measurements for each mutant.  Different categories of mutant indicated by color: defective in starch biosynthesis (red), defective in starch degradation (blue), comparative plant (green), wild types (brown), uncharacterized mutants (orange). On the left is the agglomerative clustering dendrogram obtained by the method of~\cite{PartoviNia:Davison:2012:JSS}, with the optimal clustering for this method displayed on the right.}
\label{fig:gaelledata}
\end{figure*}

In~\cite{Messerli.etal:Rapid:2007}, the metabolic pattern of 14 genetic mutants of the flowering plant \emph{Arabidopsis thaliana} are studied from measurements of 43 metabolites (mostly sugars, sugar alcohols, amino acids, and organic acids).  The 14 mutants can be described as follows: \texttt{pgm} and \texttt{isa2} are mutants defective in starch bio-synthesis; \texttt{sex1}, \texttt{sex4}, \texttt{mex1}, and \texttt{dpe2} are defective in starch degradation; \texttt{tpt} is a comparison mutant that accumulates starch as a pleiotropic effect; \texttt{WsWT}, \texttt{RLDWT}, and \texttt{ColWT} are wild-type plants; \texttt{d172}, \texttt{d263}, \texttt{ke103}, and \texttt{sex3} are uncharacterized.  Figure~\ref{fig:gaelledata} displays the raw data which consists of four replicates of metabolite measurements for each mutant, except \texttt{ColWT} which has only three.  

\subsection{Data Modeling}


The goal is to study the metabolomic characteristics of these $N = 14$ mutants via clustering.  
For this purpose, a posterior distribution $p(\cc \mid \y)$ is derived from the following hierarchical model.  A similar model has been employed by~\cite{PartoviNia:Davison:2012:JSS} for clustering on high-dimensional, small-sample datasets, and suggested for classification by~\cite{Shahbaba:Neal:BayesianHierarchicalCalssification:2007}.  The hierarchical model is
\begin{equation}\label{eq:spikeslab}
\begin{split}
y_{vc}^{\nix r} \mid \gamma_{vc}, \theta_{vc}, \eta_{vc}^{\nix} & \ind \N (\mu + \gamma_{vc}\cdot\theta_{vc}+\eta_{vc}^{\nix},\sigma^2) \\
\gamma_{vc} &\iid  \mathrm \B(p) \\
\theta_{vc} &\iid \N(0,\sigma^2_\theta) \\
\eta_{vc}^{\nix} &\iid \N(0,\sigma^2_{\eta}),
\end{split}
\end{equation}
where $\B(p)$ denotes the Bernoulli distribution with success probability $p$, and the indices $v=1,\ldots,V$, $c=1,\ldots,C$, $\nix=1,\ldots,\nobs_c$, $r=1,\ldots,R_{c\nix}$ denote, respectively, the  metabolite variables, clusters, mutant IDs within cluster, and replicate numbers.
The Bernoulli variable $\gamma_{vc}$ controls the appearance of the clustering mean $\theta_{vc}$ to adjust for noise variables. The continuous parameter $\eta_{vc}^{\nix}$ is added to account for the between-mutant error in cluster $c$.  The model parameters $\sigma^2$ and $\sigma^2_{\eta}$ are the between-replicate and between-mutant variance components, respectively, while $\sigma^2_{\theta}$ is the variance of the disappearing random mean component $\theta_{vc}$.

From model~\eqref{eq:spikeslab}, parameters $\eta_{vct}, \theta_{vc},$ and $\gamma_{vc}$ can be integrated out, resulting in a marginal likelihood mixture of two Normal distributions for each replicate:
\begin{equation}\label{eq:peb}
  \begin{aligned}
    y_{vc}^{\nix r} \sim &\, p \times \N(y_{vc}^{\nix r}; \mu, \sigma^2 + \sigma_{\eta}^2 + \sigma_{\theta}^2) \\
    & + (1-p) \times \N(y_{vc}^{\nix r}; \mu, \sigma^2 + \sigma_{\eta}^2).
    \end{aligned}
\end{equation}
In order to obtain a closed-form posterior for the cluster allocations, we employ an empirical Bayes approach.  That is, the hyperparameters $\bm{\alpha} = (\mu,\sigma^2_\eta,\sigma^2_\theta,\sigma^2,p)$ are estimated by maximizing the marginal likelihood $p(\y\mid \bm{\alpha})$ resulting from~\eqref{eq:peb}.  The estimated parameters and their standard errors are: $\hat\mu=0.083 \,(0.03),\hat\sigma^2=0.16\,(0.005),\hat\sigma^2_\theta=5.1\,(2.7),\hat\sigma^2_\eta=0.37\,(0.033)$, and $\hat p=0.034\,(0.02)$.  Upon substituting these estimates for the true parameter values, the posterior distributions $p(\gamma_{vc} \mid \y, \cc, \hat p)$, $p(\theta_{vc} \mid \y, \cc, \hat\sigma^2_\theta)$, and
$p(\eta_{vc}^{\nix} \mid \y, \cc, \hat \sigma^2_{\eta})$
are analytically tractable (they are Bernoulli, normal, and normal, respectively).  The empirical Bayes marginal likelihood is then
\begin{equation}\label{eq:eblik}
  \begin{aligned}
  p(\y \mid \cc) = 
  \prod_{vcir} & 
  \left\{\frac{p(y_{vc}^{\nix r} \mid \gamma_{vc}, \theta_{vc}, \eta_{vc}^{\nix}, \hat \mu, \hat \sigma^2) \cdot p(\gamma_{vc} \mid \hat p)}{p(\gamma_{vc} \mid \y, \cc, \hat p) \cdot p(\theta_{vc} \mid \y, \cc, \hat\sigma^2_\theta)}\right. \\
  & \left. \quad \times \frac{p(\theta_{vc} \mid \hat \sigma^2_\theta) \cdot p(\eta_{vc}^{\nix} \mid \hat \sigma^2_\eta)}{p(\eta_{vc}^{\nix} \mid \y, \cc, \hat \sigma^2_{\eta})}\right\},
  \end{aligned}
\end{equation}
where the terms in the numerator are obtained from model~\eqref{eq:spikeslab}, and those in the denominator are described above.

It now remains to specify a prior for $\cc$.  Following~\cite{Heard:Holmes:Stephens:Mosquitoes:2006}, we assume that the assignment of mutants to clusters is exchangeable.  Thus we may write
\[
\pi(\cc) =\Pr(\nobs_1,\ldots,\nobs_ C \mid C)\Pr(C),
\]
where $C$ is the number of clusters and $\nobs_c$ is the number of observations in cluster $c=1,\ldots,C$, 
such that $\nobs = \sum_{c=1}^C \nobs_c=14$ is the total number of mutants.  We employ a uniform discrete prior for the number of clusters,
\[
\Pr(C=k)=1/\nobs,\quad k=1,\ldots,\nobs,
\]
and a uniform multinomial-Dirichlet distribution for the cluster totals given the number of clusters.  This yields the prior
\begin{equation}\label{eq:prior}
\pi(\cc) \propto  \frac{(C-1)! \nobs_1!\ldots \nobs_C!}{\nobs(\nobs+C-1)!}.
\end{equation}
Combining~\eqref{eq:eblik} and~\eqref{eq:prior}, the posterior distribution on cluster allocations is
\[
p(\cc \mid \y) \propto \pi(\cc) \cdot p(\y \mid \cc).
\]

\subsection{Consensus Clustering}

\begin{figure*}[!htp]
	\centering
		\includegraphics[width=1.00\textwidth]{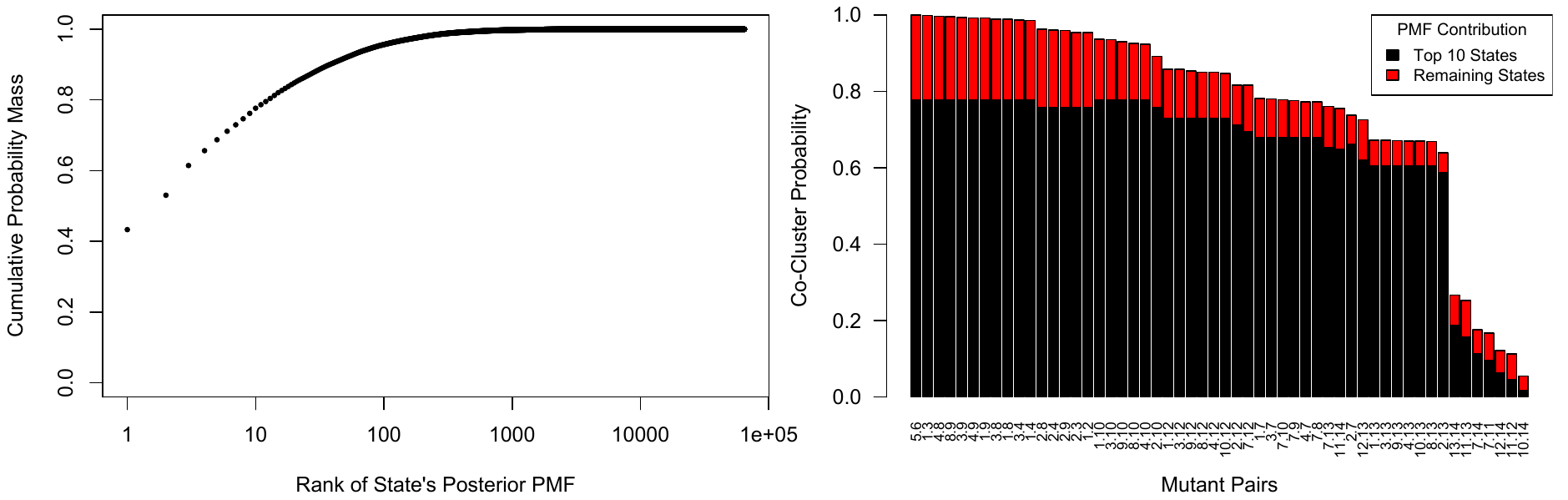}
	\caption{\emph{Left:} Cumulative PMF of cluster allocations by decreasing posterior probability.  \emph{Right:} Co-occurrence probabilities $\cp_{ij}$ for all pairs with $\cp_{ij} > .05$.  In black is the contribution of the top $K = 10$ clustering allocations, in red is that of the remainder.}
	\label{fig:truepost}
\end{figure*}

\begin{figure*}[!htp]
  \centering
  {\includegraphics[width=1.00\textwidth]{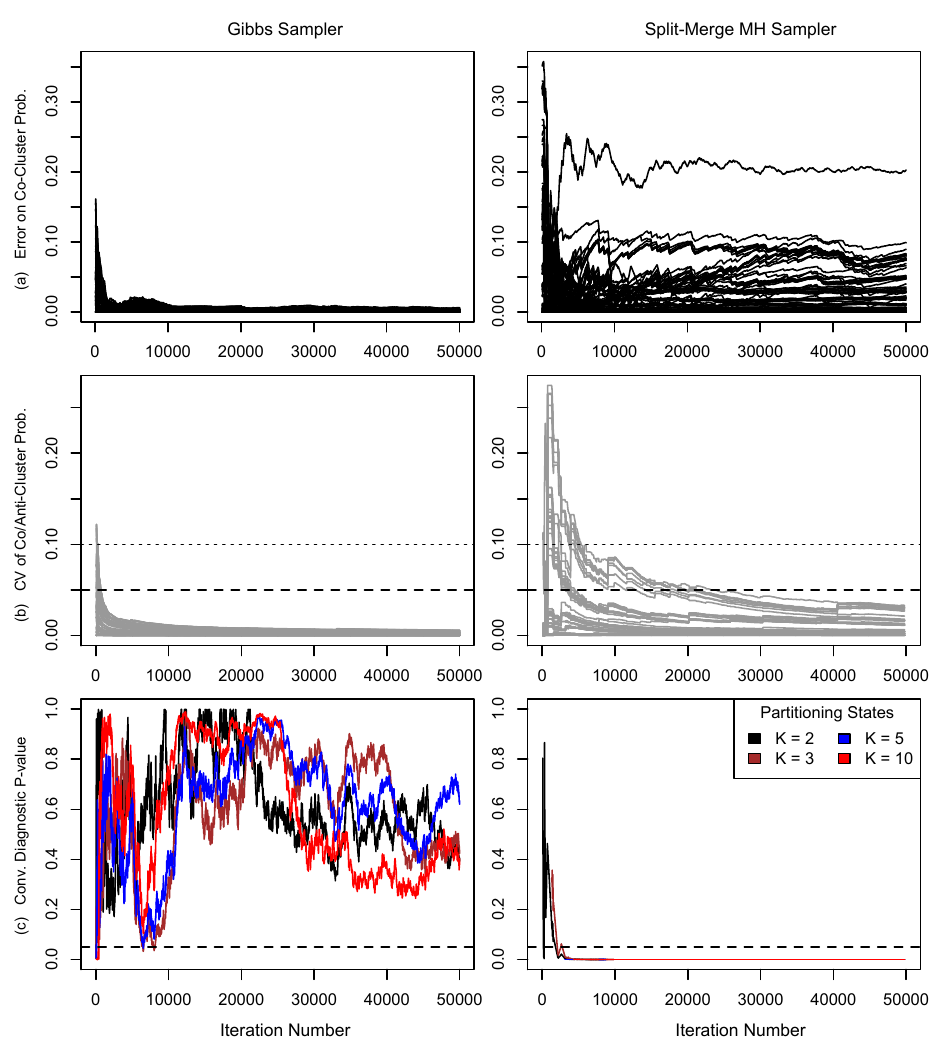}
    \phantomsubcaption\label{fig:coprob}\phantomsubcaption\label{fig:cv}\phantomsubcaption\label{fig:pval}}
  \caption{Convergence diagnosis of MCMC algorithms.  (a)  Absolute error between true pairwise co-cluster probability $\cp_{ij}$ and the regenerative sampling estimate. (b) Coefficient of variation for larger of co-cluster and anti-cluster probability estimates, $\max(\hat \cp_{ij}, 1- \hat \cp_{ij})$.  (c) P-value of the Hotelling-RS statistic $T^2_{\tau_R}$, partitioning on the $K$ most probable states.}
  \label{fig:diagcomp}
\end{figure*}

In Figure~\ref{fig:gaelledata}, the left margin displays an agglomerative clustering dendrogram produced by the Bayesian algorithm of~\cite{PartoviNia:Davison:2012:JSS}. At each step, the algorithm merges the pair of clusters which maximizes a posterior distribution similar to ours, until all mutants are in the same cluster.  The dendrogram is then cut at the maximum posterior probability on its path, resulting in the clustering allocation on the right of Figure~\ref{fig:gaelledata}.  However, agglomerative clustering is a greedy algorithm which only targets the maximum of the objective function.  

Figure~\ref{fig:truepost} displays summary information about the posterior probability on all $B(14) = 1.9\times 10^8$ cluster allocations.  Such calculations quickly become infeasible as the number of mutants increases.
The left panel of Figure~\ref{fig:truepost} displays the cumulative probability of the states, ordered by decreasing posterior probability.  Thus we can see that the maximum \emph{a posteriori} (MAP) cluster allocation is $\pi_{\textrm{MAP}} = 0.43$, as depicted by the left-most point on this graph.  About 80\% of the posterior probability is in the 10 most probable allocations, suggesting they be pooled via consensus clustering~\citep*{strehl.ghosh02,monti.etal03}.


In a recent review,~\cite{vegapons.ruizshulcloper11} describe the two main approaches to deriving a consensus clustering $\cc^\star$ from a set of candidates $\cc_1, \ldots, \cc_m$.  The first is called \emph{median partitioning}, which consists of solving for $\cc^\star = \argmax_{\cc} \sum_{t=1}^m \Gamma(\cc, \cc_t)$, where $\Gamma$ is a similarity measure between cluster allocations~\citeeg{strehl.ghosh02,filkov.skiena04}.  The second approach is based on \emph{co-occurrence}, i.e., the probability that any two observations are in the same cluster~\citeeg{fred.jain02,monti.etal03,punera.ghosh08}.  This information is contained in the consensus matrix $\bm R = [\cp_{ij}]_{1\le i,j \le N}$, of which the elements are the co-occurrence probabilities between each pair of observations $i$ and $j$.  In the Bayesian setting, each entry of $\bm R$ is defined as
\[
\cp_{ij} = \Pr(\textrm{observations $i$ and $j$ are in the same cluster} \mid \y).
\]
The consensus matrix for the mutant data is displayed in the right panel of Figure~\ref{fig:truepost}.

While the true consensus matrix corresponding to $p(\cc \mid \y)$ can be calculated exactly for $N = 14$ mutants, for larger problems it would typically be estimated by MCMC. Here we consider two sampling algorithms for the posterior distribution of cluster allocations:
\begin{enumerate}
  \item A reversible Gibbs sampler, which updates the cluster label of observations one at a time in random order;
  \item The split-merge algorithm of~\cite{Jain:Neal:SpiltMerge:2004}, which updates the cluster label of multiple observations at once.
\end{enumerate}
Both samplers were run on the mutant dataset for $n = 50,000$ iterations.  For the purpose of consensus clustering, various measures of MCMC convergence are monitored in Figure~\ref{fig:diagcomp}.

Figure~\ref{fig:coprob} displays absolute errors of the form $|\hat \cp_{ij} - \cp_{ij}|$ on the elements of the consensus  matrix, where $\hat \cp_{ij}$ is the RS estimator
\[
\hat \cp_{ij} = \frac{1}{\tau_R-1} \sum_{t=1}^{\tau_R-1} \I(\cc_t: \textrm{$i$ and $j$ in same cluster}),
\]
taken cumulatively up to the given iteration number. In this case, the pure Gibbs sampler converges to the true consensus matrix rather quickly, whereas even after $n=50,000$ iterations, the split-merge sampler estimates a good portion of the co-occurrence probabilities with 10-20\% absolute error.  This is because, for illustrative purposes, the balance between singleton and group updates in the split-merge algorithm has been deliberately tuned to achieve poor mixing. 

Figure~\ref{fig:cv} displays the convergence diagnostic of~\cite{Jones:Haran:Caffo:Neath:2006}.  That is, for each element of the consensus matrix, we compute a coefficient of variation (CV) of the form
\[
\textrm{CV}_{ij} = \frac{\textrm{se}(\hat \cp_{ij})}{\max(\hat \cp_{ij}, 1-\hat\cp_{ij})},
\]
where the standard error of the regenerative sampling estimator is given by the univariate version of~\eqref{eq:regvar}.  Note that this CV is for the larger of the co-clustering estimate $\hat \cp_{ij}$ and the anti-clustering estimate $1-\hat \cp_{ij}$. This is because the CV is a poor measure of precision when $\hat \cp_{ij} \approx 0$, whereas large values of $\max(\hat \cp_{ij},1- \hat \cp_{ij})$ are strongly informative on the co-occurrence of $i$ and $j$ (either for or against it).  While the $\textrm{CV}_{ij}$ are considerably larger for the split-merge sampler, they drop below 5\% after about $n = 20,000$ iterations.  For the purpose of estimating the consensus matrix, one might thus be misled to conclude that the sampler has converged. 

Figure~\ref{fig:pval} displays the p-value of our Hotelling-RS convergence diagnostic, using the asymptotic distribution of Corollary~\ref{thm:hot}.  Following Section~\ref{sec:prac}, we partition the sample space $\SM$ according to the $K$ most probable states, for $K = 2,3,5,10$.  In this case the convergence assessment is insensitive to the choice of $K$: while the p-values of the Gibbs sampler freely fluctuate on the $(0,1)$ interval, those of the split-merge sampler unequivocally indicate that the MCMC has not converged. 
This stands in contrast to the CV-based assessment, which cannot detect estimators that have converged to an incorrect value.

\section{Discussion}
\label{section:discuss}

We present a convergence diagnostic for MCMC on a nominal state space for which the stationary distribution is known up to a normalizing constant.  We leverage this information to check that the relative frequency of state visits is consistent with that of the equilibrium distribution.  Discrepancies between expected and observed frequencies are quantified via the p-values of the diagnostic's asymptotic distribution, which is established by Corollary~\ref{thm:hot}.

We apply the statistic to MCMC convergence assessment for Bayesian consensus clustering of $N = 14$ mutants of the plant \emph{Arabidopsis thaliana}.  Following practical recommendations in Section~\ref{sec:prac} for the implementation of our method, we find that convergence assessment is relatively insensitive to the number of top-probability states $K$ over which the sample space is partitioned.  Ostensibly, this is because most of the equilibrium distribution in our application is concentrated on a small number of states.

Here we have focused on offline convergence assessment, i.e., after running the MCMC for a predetermined number of steps.  A useful direction of future work is to evaluate convergence online, i.e., establishing at each iteration (or batch of iterations) whether another one is required.  Another line of inquiry is  extension of the diagnostic to non-conjugate Bayesian clustering models~\citep*[e.g.,][]{Tadesse:Sha:Vannucci:2005,Kim:Tadesse:Vannucci:VS-Dirichlet:2006,Jain:Neal:2007}.  
For such models the parameters cannot be integrated out, such that the (unnormalized) marginal posterior allocation probability $p(\cc \mid \y)$ is not available in closed-form -- a key requirement of the present approach.

\section*{Acknowledgements}
The authors gratefully acknowledge Professor David Stephens for comments on an earlier draft of this manuscript.


\appendix
\section{Proofs of Theorem~\ref{thm:main} and Corollary~\ref{thm:hot}}\label{appendix:asym}

First we prove a multivariate version of the regeneration sampling CLT in~\cite{Mykland:Tierney:Yu:Regeneration:1995, Hobert:Jones:Presnell:Rosenthal:2002, Jones:Haran:Caffo:Neath:2006}.  For the given state $\ii \in \SM$, the functions $s(x) = \I(x = \ii)$ and $Q(A) = \Pr(X_t \in A \mid X_{t-1} = \ii)$ trivially satisfy the minorization condition
\[
  \Pr(X_t \in A \mid X_{t-1} = x) \ge s(x) Q(A) \quad \forall\, x \in \SM, A \subseteq \SM.
\]
Then if $X_t$ is irreducible and aperiodic, it is positive Harris recurrent since $\SM$ is finite, and so for any function $g(x)$ with $\E_{\bm\Pi} \big[|g(X_1)|^{2 + \epsilon}\big] < \infty$ for some $\epsilon > 0$, Theorem 2 of~\cite{Hobert:Jones:Presnell:Rosenthal:2002} establishes the regenerative sampling CLT
\[
  \sqrt{R} \big( \bar{g}_{\tau_R} - \E_{\bm\Pi} [g(X_1)] \big) \Dto \N_1(0, \sigma^2_g) \as R \to \infty,
\]
and consistency of the variance estimator
\[
  \hat \sigma^2_{\tau_R} = \frac{1}{R\bar N^2} \sum_{r=1}^R (s_r - N_r \bar g_{\tau_R})^2 \to \sigma^2_g,
\]
where $s_r = \sum_{t=\tau_{r-1}}^{\tau_r-1} g(X_t)$.  In particular, this holds for $g(x) = \bm a' \bm g(x)$, where $\bm a$ is an arbitrary vector in $\mathbb R^K$ and $\bm g(x) = \big(\I(x \in \II_1)/q_1, \ldots, \I(x \in \II_K/q_K)\big)$ as defined in the statement of Theorem~\ref{thm:main}.  Since a univariate CLT holds for any linear combination of $\bar \g_{\tau_R}$, by the Cram\'er-Wold device we have the multivariate CLT
\[
  \sqrt{R} \big( \bar{\g}_{\tau_R} - \E_{\bm\Pi} [\g(X_1)] \big) \Dto \N_K(0, \SS_g) \as R \to \infty.
\]
Recall that $\E_{\bm\Pi} [\g(X_1)] = \bm 1_K Z^{-1}$, where the value of the normalizing constant $Z$ is unknown.  Then for $\bm A : \mathbb R^{K\times K} \to \mathbb R^{K\times K}$ defined by
\[
  \bm A(\SS) = \SS^{-1/2} - \SS^{-1/2}\bm 1_K(\bm 1_K' \SS^{-1} \bm 1_K)^{-1}\bm 1_K'\SS^{-1}
\]
we have
\begin{multline*}
  \bm A(\SS_\g) \cdot \sqrt R (\bar \g_{\tau_R} - \bm 1_K Z^{-1})  = \\
  \sqrt R \cdot \bm A(\SS_\g) \bar \g_{\tau_R} 
  \Dto \N_K\big(\bm 0, \bm B(\SS_\g)\big),
\end{multline*}
where
\[
  \bm B(\SS) = \bm I_K - \SS^{-1/2}\bm 1_K(\bm 1_K'\SS^{-1}\bm 1_K)^{-1} \SS^{-1/2}.
\]
A straightforward calculation shows that $\bm B(\SS)$ is idempotent with rank $K-1$, such that
\begin{equation}\label{eq:achi}
R \cdot \big[\bm A(\SS_\g) \bar \g_{\tau_R}\big]'\big[\bm A(\SS_\g) \bar \g_{\tau_R}\big] \Dto \chi^2_{(K-1)}.
\end{equation}
Moreover, for any consistent estimator $\hat \SS_R \to \SS_\g$, we have $\bm A(\hat \SS_R) \to \bm A(\SS_\g)$, such that the multivariate version of Slutsky's theorem gives $\sqrt R \cdot \bm A(\SS_R) \bar \g_{\tau_R} \Dto \N_K\big(\bm 0, \bm B(\SS_\g)\big)$, such that~\eqref{eq:achi} holds with $\SS_R$ in place of $\SS_\g$.

Again, a straightforward calculation gives
\begin{align*}
  T^2_{\tau_R}  & := R \cdot \big[\bm A(\SS_{\tau_R}) \bar \g_{\tau_R}\big]'\big[\bm A(\SS_{\tau_R}) \bar \g_{\tau_R}\big] \\
   & \phantom{:}= R \cdot(\bar \g_{\tau_R} - \hat Z_{\tau_R}^{-1})'\hat \SS_{\tau_R}^{-1}(\bar \g_{\tau_R} - \hat Z_{\tau_R}^{-1}),
\end{align*}
where $\hat Z_{\tau_R}^{-1} = \bm w(\hat \SS_{\tau_R})'\bar \g_{\tau_R}$.
 It now remains to show that $\hat \SS_{\tau_R}$ in~\eqref{eq:regvar} is a consistent estimator of $\SS_\g$.  For any vector $\bm a \in \mathbb R^K$ and $g(x) = \bm a' \g(x)$, note that the consistent estimator of $\sigma^2_g = \bm a' \SS_\g \bm a = \lim_{R\to\infty} R \var(g_{\tau_R})$ defined by~\cite{Hobert:Jones:Presnell:Rosenthal:2002} is given by $\hat \sigma^2_{\tau_R} = \bm a' \hat \SS_{\tau_R} \bm a$.  Thus by picking $\bm a = \bm e_i$, the $i$th standard basis vector of $\mathbb R^K$, we find that $[\hat \SS_{\tau_R}]_{ii} \to [\SS_\g]_{ii}$.  
For the off diagonal elements, let $\bm a_1 = \bm e_i + \bm e_j$ and $\bm a_2 = \bm e_i - \bm e_j$.  Then
\begin{align*}
  \tfrac 1 4 \big(\bm a_1' \SS_\g \bm a_1 - \bm a_2' \SS_\g \bm a_2\big) & = [\SS_\g]_{ij}, \\
  \tfrac 1 4 \big(\bm a_1' \hat\SS_{\tau_R} \bm a_1 - \bm a_2' \hat \SS_{\tau_R} \bm a_2\big) & = [\hat \SS_{\tau_R}]_{ij},
\end{align*}
and convergence of linear combinations of random sequences converging to constants implies that $[\hat \SS_{\tau_R}]_{ij} \to [\SS_\g]_{ij}$.  Finally, elementwise convergence implies that $\hat \SS_{\tau_R} \to \SS_\g$ by the equivalence of matrix norms.  Thus we have
\[
T^2_{\tau_R} = R \cdot \big[\bm A(\SS_{\tau_R}) \bar \g_{\tau_R}\big]'\big[\bm A(\SS_{\tau_R}) \bar \g_{\tau_R}\big] \Dto \chi^2_{(K-1)}.
\]
\hfill $\qedsymbol$


\bibliographystyle{stdref}
\bibliography{convergence-ref}

\end{document}